\documentclass[useAMS,usenatbib]{mnras}

\usepackage{tabularx}
\usepackage{multirow}
\usepackage{url}
\usepackage{amsmath}	
\usepackage{amssymb}	
\usepackage{aas_macros}
\usepackage{graphicx}
\usepackage{ulem,soul}

\usepackage{times,txfonts}
\usepackage{bm}

\usepackage[T1]{fontenc}
\usepackage{ae,aecompl}
\newcommand\lb{\langle}
\newcommand\rb{\rangle}

\title[Transport in extended shearing boxes]{Transport properties of Keplerian flows in extended local domains with no imposed field}

\author[F. Nauman and M. Pessah]{Farrukh Nauman$^{1}$\thanks{E-mail: nauman@nbi.ku.dk}, 
	Martin E. Pessah$^{1}$\thanks{E-mail: mpessah@nbi.dk}\\
	$^{1}$Niels Bohr International Academy, The Niels Bohr Institute, Blegdamsvej 17, DK-2100, Copenhagen \O, Denmark.
}

\pubyear{2017}

\begin{document}
	\date{\today}
	
	\pagerange{\pageref{firstpage}--\pageref{lastpage}}
	
	\maketitle
	
	\label{firstpage}

\begin{abstract}
	We compare transport statistics of elongated incompressible shearing boxes for different Reynolds and magnetic Prandtl numbers, $Re$ and $Pm$, and aspect ratios, $L_z/L_x$. We find that at fixed aspect ratio $L_z/L_x=4$ and $Re = 10,000$, the turbulent stresses for $Pm \lesssim 1$ do not show considerable variation and follow a power law $\sim Pm^{3/2}$ for $Pm > 1$. This is qualitatively consistent with previous results based on net imposed flux and small box $L_z/L_x \sim 1$ simulations but the power law exponent is different. The saturated level of stresses, the ratio of Maxwell stress to the magnetic energy and Reynolds to Maxwell stress ratio are roughly invariant as $L_z/L_x$ is increased. For cases where the boxes are elongated in both the azimuth and vertical direction, the transport coefficient $\alpha \in [0.1,1.0]$ that is $10-100$ times larger than the case with $L_y/L_x = 2$ and large $L_z/L_x$. Overall, our results suggest that the saturated state of turbulence is sensitive to both dissipation coefficients and aspect ratio (both $L_z/L_x$, $L_y/L_x$) motivating further work on this problem. 
\end{abstract}

	
\maketitle

\section{Introduction}  \label{sec:intro}
Accretion flows exist in a variety of astrophysical systems. The accreting fluid would have a much longer lifetime if molecular viscosity was the sole source of angular momentum transport. For this reason, a number of possible alternative transport mechanisms have been investigated in recent years ranging from purely hydrodynamic sources to magnetohydrodynamic sources. Evaluating each of these mechanisms in detail and understanding their observational implications is a subject of ongoing research \citep{bh1998,fromang2017}.

Magnetized Keplerian flows are linearly unstable to the magnetorotational instability (MRI) if a weak external magnetic field is present (\citealt{1959velikhov, 1960PNAS...46..253C, 1991ApJ...376..214B}). The MRI has been studied extensively in local shearing box simulations \citep{hawley1995, brandenburg1995, fromang2007}, global disk simulations \citep{flock2012, parkin1, suzuki2014, zhustone} and Taylor-Couette flow simulations \citep{wei2016,guseva2017apj,guseva2017prl}. Despite more than two decades of numerical work on the problem, the issue of convergence is still unresolved  with work focusing on imposed field, the presence of  dissipation coefficients, density stratification due to gravity \citep{fromang2007convergence, davis2010, hawley2011, bodo2014, meheut, ryan2017}.

If the external flux is removed in a magnetized Keplerian flow, then the fluid is no longer linearly unstable to MRI but is yet observed to reach a nonlinear steady state with significant transport \citep{hawley1996, fromang2007, lesurogilvie2008, guseva2017prl}. In the literature, this case is sometimes referred to as the `dynamo' case but we do not use such terminology here because in a broad sense, fluids linearly unstable to MRI are also an example of a dynamo since they lead to the generation and sustenance of magnetic fields \citep{gressel2015}. The `zero flux' case has been a particular focus of numerical studies trying to seek convergence since it was realized that the increase in resolution leads to a decrease in transport \citep{pessah2007, fromang2007convergence}. 

The role of Reynolds number as an important parameter in shear flows is well known since the original work of Reynolds in 1883. The realization that the aspect ratio might also play an important role in determining the nonlinear state of a fluid came much later \citep{cross1993, philip2011}. Moreover, recent numerical and experimental work on shear flows suggests that small domains might suffer from finite size effects at transition \citep{lemoult2016NatPh} and that the transition to turbulence in shear flows perhaps belongs to the directed percolation universality class \citep{pomeau1986}.

Until recently, Keplerian flows without a net imposed flux were thought to be stable for magnetic Prandtl number at and below unity \citep{fromang2007}. However, \citet{naumanpessah2016} (Paper I) showed that if a large aspect ratio ($L_z/L_x \ge 4$) is used, then flows can still reach a nonlinear steady state with magnetic Prandtl number below unity. The transport properties of such elongated boxes was not addressed in that paper. The focus of the current paper is to explore the dependence of turbulent transport on box size, Reynolds number and magnetic Prandtl number.


That the aspect ratio plays an important role has gained further support from recent work by \cite{shi2016} and \cite{walker2017} with the observation that larger aspect ratios allow for local non-zero toroidal magnetic flux. Using ideal compressible MHD simulations, \cite{shi2016} showed not only that the turbulent stresses in boxes with $L_z/L_x > 2.5$ converged with respect to increasing resolution but also that the saturated level of stresses was insensitive to further increase in the aspect ratio. The bulk of their work was using ideal MHD and while \cite{shi2016} did some numerical simulations with explicit dissipation coefficients, a more comprehensive survey of the effects of aspect ratio in the presence of dissipation has yet to be carried out.

The goal of this paper is to study the transport properties as a function of aspect ratio and dissipation coefficients. The organization of this paper is as follows: Section \ref{sec:methods} describes the numerical setup. Section \ref{sec:largeaspect} reports on transport properties of large aspect ratio systems as a function of $Re$, $Pm$ and $L_z$. We then present resolution tests in section \ref{sec:resolution}. In section \ref{sec:extended}, we describe the results from simulations where the domain is extended in both `$y$' and `$z$'. We conclude in section \ref{sec:conclusions}.

\section{Numerical Setup} \label{sec:methods}
Using the publicly available pseudospectral code \textsc{snoopy} \footnote{\url{http://ipag.osug.fr/~lesurg/snoopy.html}} (\citealt{lesur2007}), we solved the incompressible MHD simulations in the shearing box framework:
\begin{gather}
\frac{\partial \bm{v}}{\partial t} + V_{sh} \frac{\partial \bm{v}}{\partial y} + \nabla \cdot \bm{T} = 2 \Omega v_y \bm{e}_x - (2 - q) \Omega v_x \bm{e}_y + \nu \nabla^2 \bm{v}, \label{eq:NS} \\
\frac{\partial \bm{b}}{\partial t} = \nabla \times ({\bm V}_{\text{sh}} \times \bm{b}) + \nabla \times (\bm{v} \times \bm{b}) + \eta \nabla^2 \bm{b}, \label{eq:induc}\\
\nabla \cdot \bm{v} = 0, \\
\nabla \cdot \bm{b} = 0, 
\end{gather}
where $\bm{v}$ and $\bm{b}$ are the velocity and magnetic field respectively. Here $\bm{T}$ is a stress tensor given by
\begin{equation}
\bm{T} = (p + b^2/2) \bm{I} + \bm{v} \bm{v} - \bm{b}\bm{b},
\end{equation}
where $\bm{I}$ is the identity matrix and $p$ is thermal pressure. 

We impose no external magnetic flux. The magnetic field is initialized with a sinusoidal profile: ${\bm B}_{\text{ini}} = B_0 \sin(k_x x)\, {\bm e}_z$. The background velocity profile is given by ${\bm V}_{\text{sh}} = - Sx\, {\bm e}_y$, where $S=q\Omega=1$ ($q=-d\ln\Omega/d\ln r=3/2$ for Keplerian shear) is the shear parameter and $\Omega$ is the angular frequency. Since this state is linearly stable, we have to apply large amplitude perturbations to trigger nontrivial dynamics. The perturbations are of magnitude $L S$ applied to large scale velocity modes. The time unit in our simulations is the shear time, $1/S$ ($\sim 1/10$ orbits where $1$ orbit = $2\upi/\Omega$). The magnetic field has the units of the Alfven speed and $\umu_0=1$,$\rho_o=1$. The Reynolds \ and magnetic Reynolds numbers are $Re = SL_x^2/\nu$ and $Rm = SL_x^2/\eta$, respectively (with $L_x = 1$ and $Pm = Rm/Re$). We point out that the definition of $Re$ and $Rm$ we use is different from some of the previous works we refer to: compressible studies typically take the pressure scale height, $H$, as the length scale instead of $L_x$ while $L_z$ has been used by \citet{lesur2007} for incompressible studies.

\section{Large aspect ratio: $L_z/L_x \ge 4$} \label{sec:largeaspect}
We describe the different transport properties in this section. Note that, unless otherwise indicated, all quantities are volume averaged over the entire domain and time averaged over $500-10000$ $S^{-1}$ (or roughly $50-1000$ orbits except for $L_z=32$ which was only averaged over $500-1000$ $S^{-1}$). In the following, we compare our results with the only other study that focused on aspect ratio dependence of transport coefficients, \cite{shi2016}. It is important to mention two key differences in their simulations and ours: \cite{shi2016} use ideal MHD for most of their work and solve compressible MHD equations. We, on the other hand, have run incompressible simulations with explicit dissipation coefficients. Moreover, we focus our attention to studying the effect of varying aspect ratios and dissipation coefficient at the same resolution while \cite{shi2016} focused on resolution studies for different aspect ratios and had a small fraction of runs including explicit dissipation. A resolution study is presented in section \ref{sec:resolution} that does not show drastic differences between the resolution adapted for most of this work ($64/L_x$) and higher resolutions ($128/L_x, 256/L_x$). We also point out that the runs $L_z/L_x=16,32$ are new to this paper and were not part of the study in Paper I. 

\subsection{Fixed $Re=10,000, Pm=1$, variable $L_z$}


\begin{figure}
	\centering
	\includegraphics[width=0.475\textwidth]{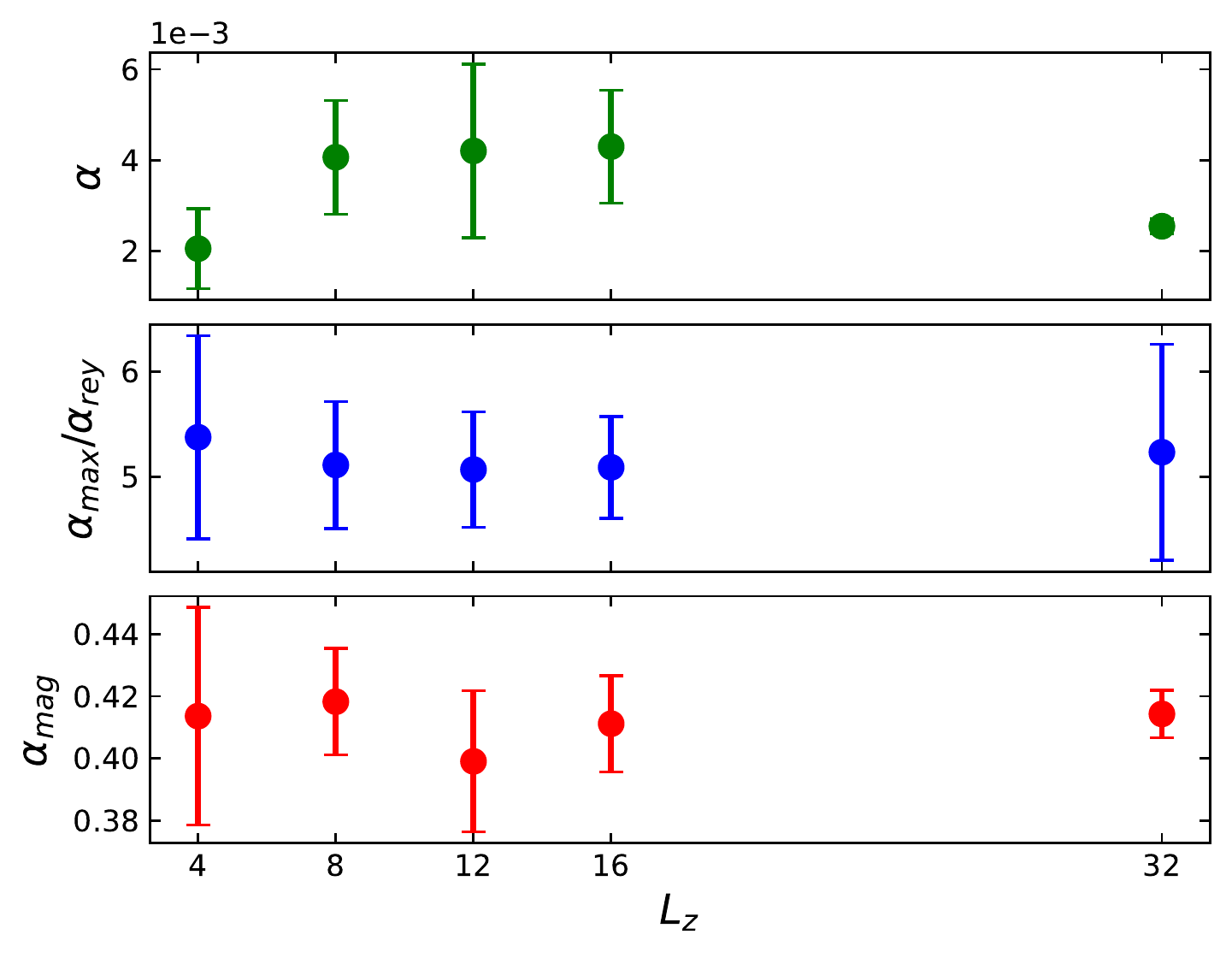}
	\caption{The volume averaged stress, $\alpha = \langle v_x v_y - b_x b_y\rangle/S^2L_x^2$ that was further averaged over time from $t=500-10000$ $S^{-1}$ (except $L_z=32$ that was averaged in the range $250-1000$ $S^{-1}$) at $Re=10,000, Pm=1$ as a function of different $L_z$. \textit{Top}: The stresses at $L_z/L_x \geq 8$ are all higher than the $L_z/L_x=4$ case. \textit{Middle}: The ratio of Maxwell to Reynolds stress, $\langle -b_xb_y/v_xv_y\rangle$ remains roughly constant for different $L_z/L_x$. \textit{Bottom}: The ratio of the Maxwell stress to magnetic energy, $\alpha_{\text{mag}}=\langle -b_xb_y/b^2\rangle$ is also rather insensitive to the aspect ratio for the cases considered here. The first and second results are in agreement with the work of \protect\cite{shi2016} while the lack of sensitivity of $\alpha_{\text{mag}}$ to aspect ratio is not.}
	\label{fig:alphavsLz}
\end{figure}

\cite{shi2016} demonstrated using compressible ideal MHD simulations that the stresses converge with respect to resolution for $L_z/L_x \geq 2.5$ and that this saturation level was nearly independent of the aspect ratio beyond $L_z/L_x=2.5$. In our incompressible simulations, we could not find any sustained turbulence for $L_z/L_x < 4$ at $Pm = 1$ so we are unable to confirm if $L_z/L_x=2.5$ defines a threshold for convergence. In fig. ~\ref{fig:alphavsLz}, we plot the volume averaged stress, $\alpha = \langle v_x v_y - b_x b_y\rangle/S^2L_x^2$ as a function of $L_z$, which seems to be slightly sensitive to the aspect ratio but does not show great variation going from $L_z/L_x=4$ to $32$. In the middle panel, we plot the Maxwell to Reynolds stress ratio, which seems to converge between $5-6$. Both this and the $\alpha$ convergence are consistent with \cite{shi2016} results (see their figures 6 and 7). The bottom panel shows that $\alpha_{\text{mag}} = \langle -b_xb_y/b^2\rangle$ remains approximately invariant $0.40-0.43$ as the aspect ratio is increased. This differs from the behavior reported in fig. 8 of \cite{shi2016}, where they found that the $\alpha_{\text{mag}}$ decreased with increasing aspect ratio as $(L_z/L_x)^{-1/2}$. 

\subsection{$Pm$ dependence}
The dependence of $\alpha$ on dissipation coefficients is of interest since accretion disks come in a wide variety: protoplanetary disks that have $Pm \ll 1$ and active galactic nuclei with $Pm \sim 1$. Moreover, laboratory experiments of Taylor-Couette flow are typically done with liquid metal that have $Pm \in [10^{-7},10^{-5}]$. This makes detection of MRI in the lab a particularly challenging problem since MRI with a vertical field is hard to trigger for such low Pm. At low $Pm$, a net azimuthal or helical field might make it easier to trigger turbulence in magnetized Keplerian flows ( \citealt{guseva2017apj}; see \citealt{rudiger2017} for a review of laboratory MRI). 
\begin{figure}
	\centering
	\includegraphics[width=0.475\textwidth]{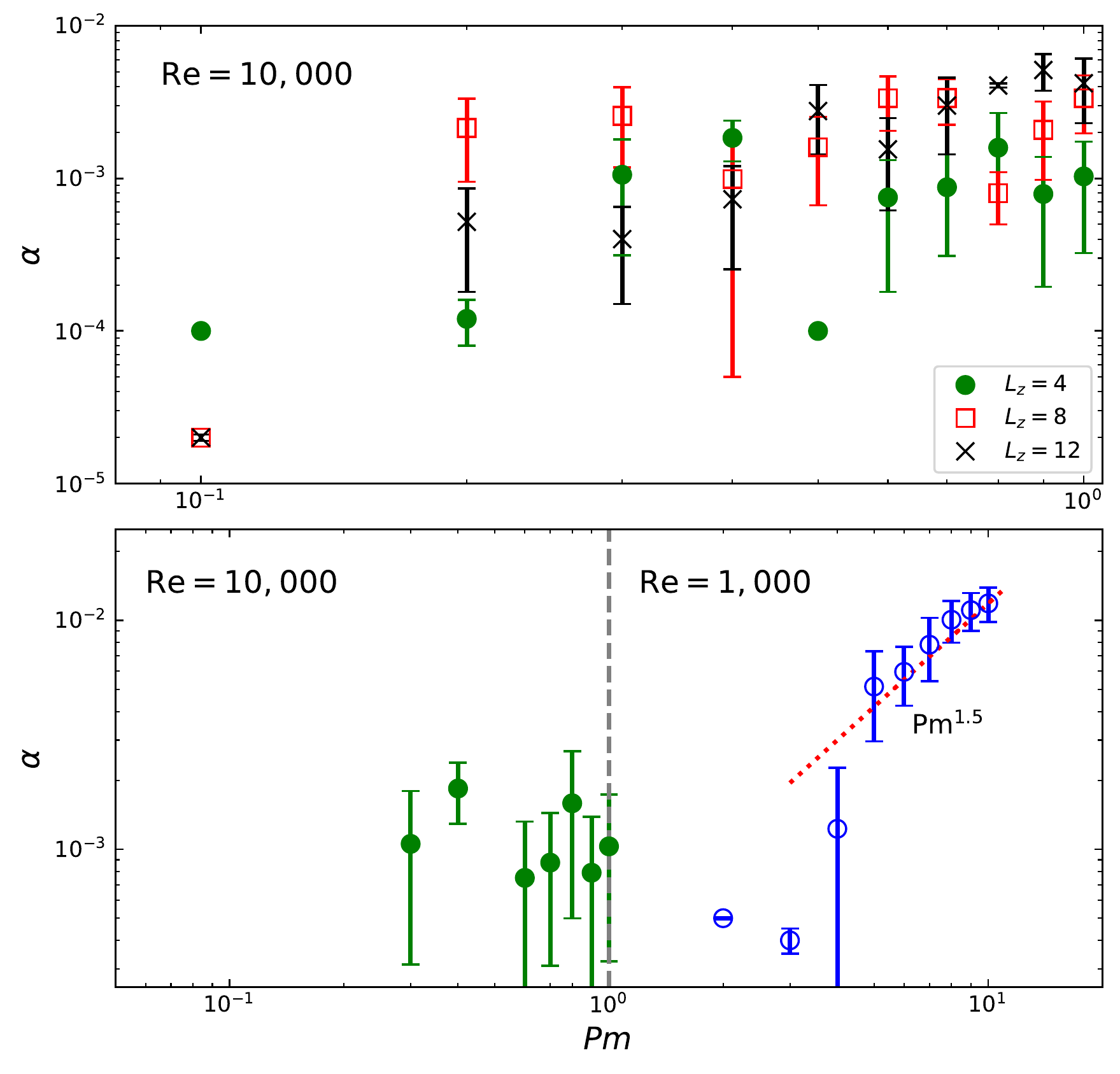}
	\caption{Volume averaged stress, $\alpha = \langle v_x v_y - b_x b_y\rangle/L_x^2S^2$, on a log-log plot further averaged over time from $t=250-1000$ $S^{-1}$ (except for cases that decayed before that). \textit{Top}: Stress for $Pm<1$ at $Re=10,000$ as a function of different aspect ratios $L_z/L_x$. The $L_z/L_x=8$ (red squares) and $L_z/L_x=12$ (black crosses) show variations up to an order of magnitude while $L_z/L_x=4$ is more or less constant ($Pm=0.1,0.2,0.5$ are nearly zero for this case). \textit{Bottom}: At fixed $L_z=4$, the stress seems to be insensitive to $Pm$ for $Pm<1$ (filled green circles) while it follows a $\alpha \sim Pm^{1.5}$ scaling for $Pm>1$ (hollow blue circles). This is qualitatively consistent with previous work done with imposed fields \protect\citep{meheut}. In the bottom plot, data points for $Pm = 0.1,0.2,0.5$ are ignored.}
	\label{fig:alphavsPm1}	
\end{figure}

In Paper I, it was shown that turbulence could only sustain if $L_z/L_x \geq 4$ in the $Pm < 1$ regime with no imposed field. Previous studies addressing the dependence of transport coefficients on dissipation were confined to small aspect ratio shearing boxes and a net magnetic flux \citep{lesur2007, fromang2007, meheut}. These previous works found that the $Pm$ dependence of $\alpha$ was very weak for $Pm<1$, while it followed a power law for $Pm>1$. \cite{guseva2017apj} found qualitative similar results for magnetized Taylor Couette flow. Our results are mixed (fig. ~\ref{fig:alphavsPm1}): for $L_z/L_x=4$ stress is insensitive to $Pm$ at low $Pm$ and has power law dependence $\alpha\sim Pm^{3/2}$ for higher $Pm$ (bottom panel). However, $L_z/L_x=8$ shows variations up to a factor of 3 for $Pm<1$ (top panel). Note, however, that a single run at each ($L_z/L_x$, $Pm$) is potentially misleading and ideally one would want several runs initiated with different initial conditions (each evolved for a long time ($\gg 100$ orbits)).

\subsection{Role of mean fields}
\begin{figure*}
	\centering
	\includegraphics[width=0.95\textwidth]{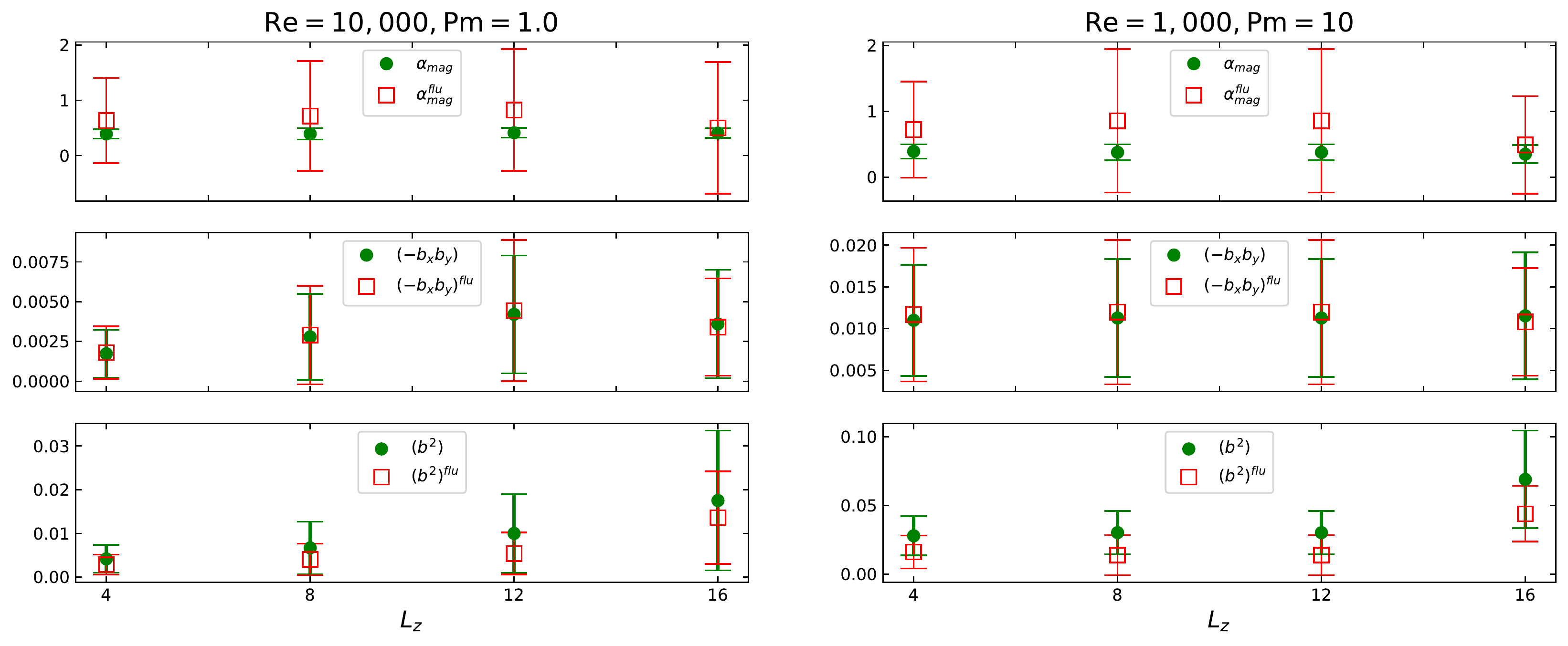}
	\caption{The horizontally ($xy$) averaged mean field and the corresponding fluctuating quantities that were further averaged over $t,z$ for $L_z=4,8,12,16$ at $Re=1,000, Pm=10$ and $Re=10,000, Pm=1$. These quantities have been time averaged over $500-10000$ $S^{-1}$. \textit{Top}: Neither $\alpha_{\text{mag}}$ nor the corresponding fluctuating quantity change much with either the vertical aspect ratio or the two different $Pm$. \textit{Middle}: The total and fluctuating Maxwell stresses seem to be insensitive to box size but they are sensitive to $Pm$ as we showed in the previous section. \textit{Bottom}: The difference between the total and fluctuating magnetic energy is within a factor of 2. For both $Pm=1$ (left) and $Pm=10$ (right), the total and fluctuating magnetic energy grow with aspect ratio.}
	\label{fig:meanvsfluct}
\end{figure*}

We define the mean field $\overline{Q}$ as:
\begin{equation}
	\overline{Q} = \frac{1}{L_xL_y}\int Q dx dy.
	\label{eq:mean}
\end{equation}

This definition of mean field is not universal but it makes comparison with \cite{shi2016} easier since they also employed a horizontal average. The disadvantage of using a horizontal average is that the vertical small scales (close to dissipation) are also considered part of the `mean'. In fig. \ref{fig:meanvsfluct}, we plot the total and fluctuating $\alpha_{\text{mag}}$, Maxwell stress and the magnetic energy. The fluctuations in magnetic field are defined through: ${\bm B} =  \overline{\bm B} + {\bm b'}$ while $\alpha^{\text{flu}}_{\text{mag}}$ is defined as:
\begin{equation}
	\alpha^{\text{flu}}_{\text{mag}} = \frac{-\overline{b'_xb'_y}}{\overline{b'^2}}
	\label{eq:alphamagflu}
\end{equation}
where ${\bm b'}$ is the fluctuating magnetic field. We do not observe clear differences in convergence between the two sets of quantities, $\alpha_{\text{mag}}$ and $\alpha^{\text{flu}}_{\text{mag}}$ (see top panel of fig. \ref{fig:meanvsfluct}). This is in contrast to \protect\cite{shi2016} who found that it is the $\alpha^{\text{flu}}_{\text{mag}}$ that is invariant with vertical aspect ratio but the total $\alpha_{\text{mag}}$ decreases. Furthermore, \cite{shi2016} observed that while the total stress $\alpha$ converged for aspect ratios $L_z/L_x > 2.5$, the total $\alpha_{\text{mag}}$ decreased as $(L_z/L_x)^{-1/2}$ with increasing aspect ratio. For the stresses, no clear difference in trends seems to exist as a function of aspect ratios between the total or the fluctuating quantities (see also fig. ~ \ref{fig:alphavsLz}). The mean $\alpha^{\text{mean}}_{\text{mag}} = -\overline{B_x} \overline{B_y}/\overline{B}^2 \sim 10^{-2}$ for all the aspect ratios considered here and is considerably smaller than the fluctuating contribution in agreement with \cite{BNreview,shi2016}. However, the total and fluctuating magnetic energies (bottom panel on left and right) seem to increase with aspect ratio.

\section{Resolution tests} \label{sec:resolution}
\begin{figure}
	\centering
	\includegraphics[width=0.475\textwidth]{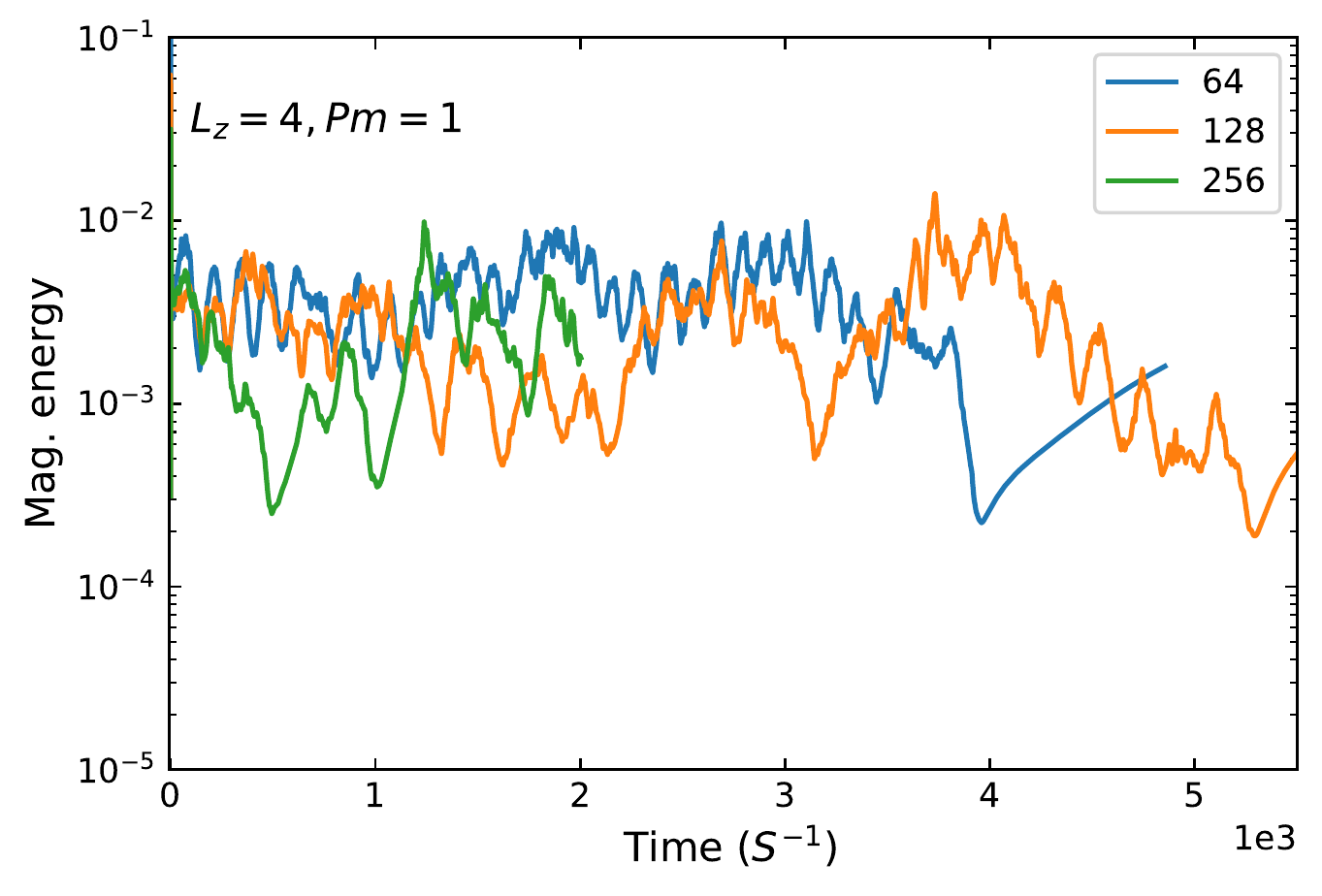}
	\includegraphics[width=0.475\textwidth]{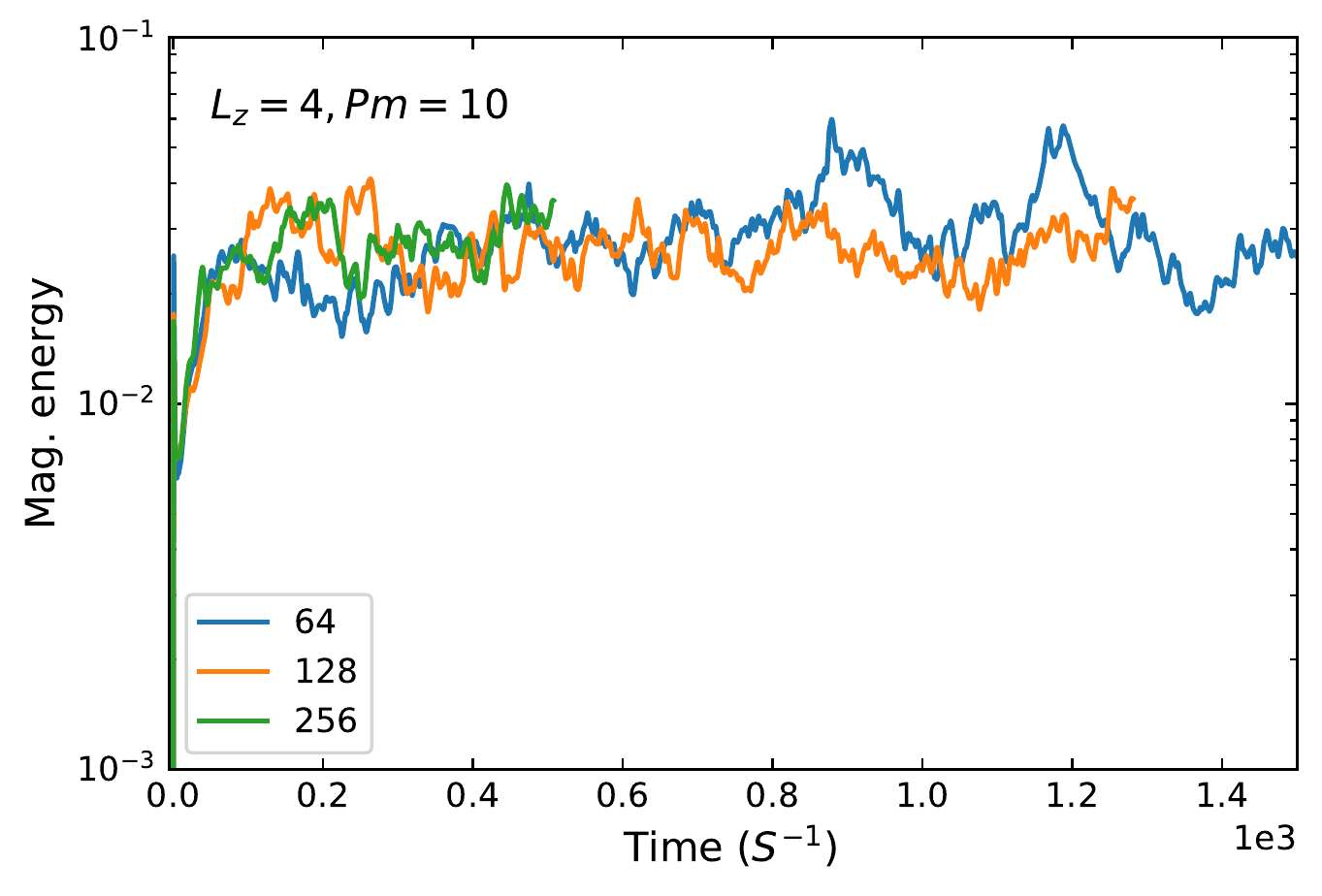}
	\caption{Time history plotted for different resolutions at $L_z=4$, $Pm=1$ (top) $Pm=10$ (bottom). \textit{Top}: The decay time scale as well as behavior at the early stages is sensitive to resolution. The time averaged amplitude of the stresses in the saturated state averaged over $500-2000$ $S^{-1}$ is $0.002,0.0008,0.0009$ for resolutions $64/L_x,128/L_x,256/L_x$. \textit{Bottom}: No strong variations among the three runs seem to exist ($\alpha \sim 0.012$ for all three runs).}
	\label{fig:restest1}
\end{figure}

The results reported in this paper describe the statistics of quadratic quantities (stresses and energies) as a function of dissipation coefficients and aspect ratios. One might ask if these results are sensitive to the numerical resolution employed ($64/L_x$). We first note that pseudo-spectral methods (as employed in \textsc{snoopy}) are more accurate than finite difference methods: a rule of thumb is that second order (central difference) finite difference schemes require twice as much resolution as a pseudo-spectral scheme to achieve the same level of accuracy (\cite{moin1998}). Secondly, higher order moments (for example, fourth order structure functions) require higher resolution to be properly resolved but the resolution requirements are not as stringent on quadratic quantities like energies or stresses (\cite{yakhot2005,donzis2008}).

\begin{figure}
	\centering
	\includegraphics[width=0.475\textwidth]{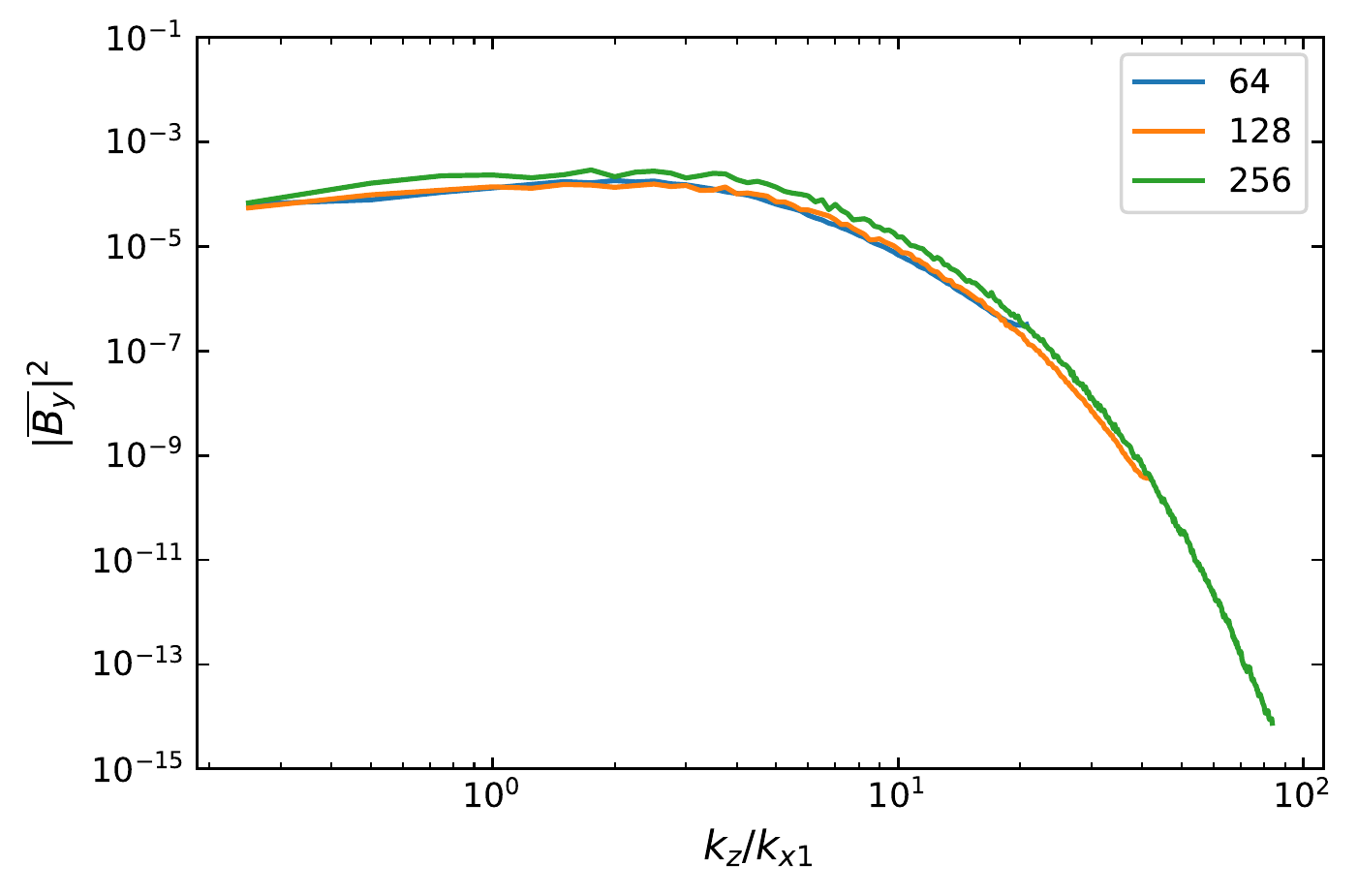}
	\includegraphics[width=0.475\textwidth]{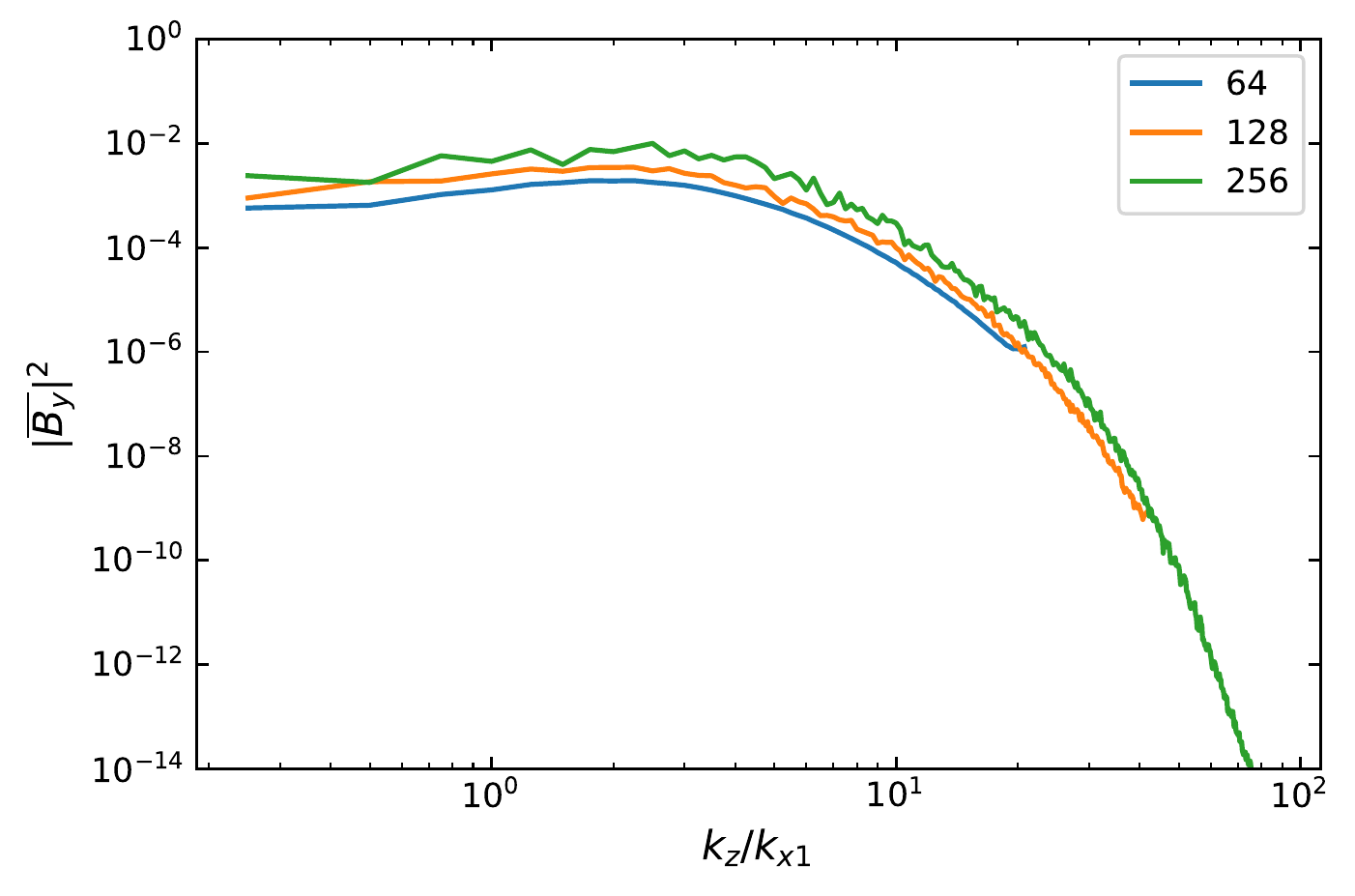}
	\caption{Power spectrum for the horizontally averaged $\overline{B_y}$ as a function of vertical wavemode for different resolutions at $L_z=4$, $Pm=1$ (top) $Pm=10$ (bottom). The spectrum is normalized such that $\int |\overline{B_y}|^2 dk_z = \langle B_y^2 \rangle$. The vertical wavemode on the horizontal axis is normalized to $k_{x1}=2\pi/L_x$. The overall shape of the spectra is very similar for all the 3 resolutions for both the top and bottom set of runs. We note that the temporal average is over fewer snapshots for higher resolution runs since we only stored snapshots every $10$ $S^{-1}$ (for $256\times 256\times 1024$) as opposed to $1$ $S^{-1}$ for $64/L_x,128/L_x$.}
	\label{fig:restest2}
\end{figure}

Nonetheless, it is useful to do a resolution study to check if there are any drastic differences between the resolutions employed in this paper and higher resolutions. For this purpose, for two runs with $L_x=1,L_y=2,L_z=4,Rm=10000$ having different $Pm$ ($=1,10$), we study (i) the time history of volume averaged stresses in fig. \ref{fig:restest1}; and (ii) the power spectrum of $\langle v_x\rangle$ as a function of $k_z$ in fig. \ref{fig:restest2}. The plots in fig. \ref{fig:restest2} were calculated in postprocessing and were consequently time averaged over significantly smaller duration as well as snapshots as the size of each snapshot significantly increases with resolution. We do not notice any noticeable difference between the lower resolution $64/L_x$ and higher resolutions $128/L_x,256/L_x$ (fig. \ref{fig:restest2}).

Higher resolution runs are computationally very expensive. For instance, the run at $256/L_x$ for $L_z=4, Re=Rm=10000$ costed $\sim 1.5\times10^5$ CPU hours for just $2000 S^{-1}$ while all the runs combined in fig. \ref{fig:alphavsLz} took less than $10^5$ CPU hours. These runs become even more expensive as the energies increase with $Pm$. The $L_z=4, Re=1000, Pm=10$ run took $10^5$ CPU hours for just $500 S^{-1}$, which means that it would take $2\times10^6$ CPU hours to evolve this run to $10000 S^{-1}$.

\section{Extended y and z domains} \label{sec:extended}
\begin{figure}
	\centering
	\includegraphics[width=0.475\textwidth]{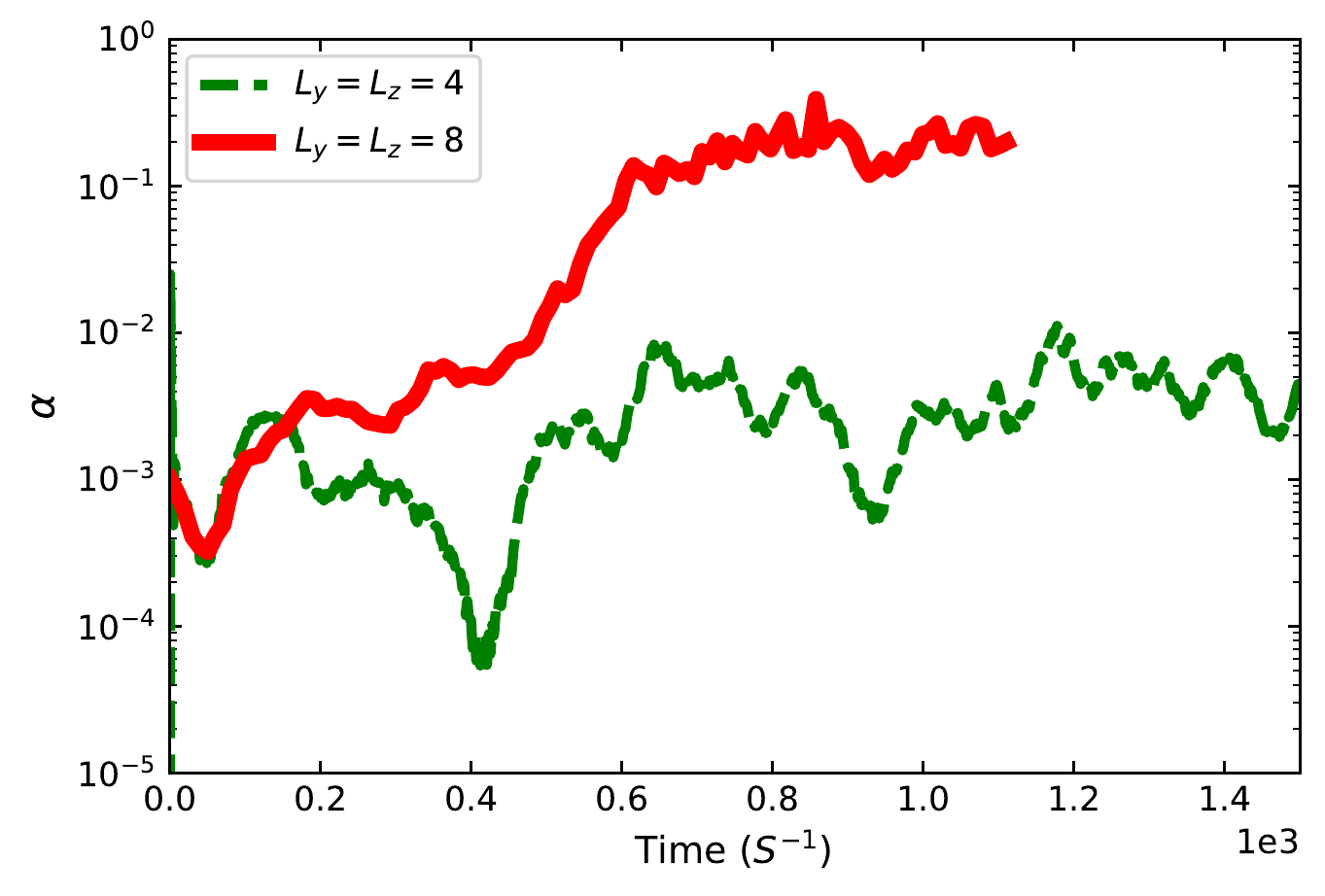}
	\label{fig:alphaLyLz}
	\caption{Volume averaged stress, $\alpha = \langle v_x v_y - b_x b_y\rangle/L_x^2S^2$ as a function of time for $L_y=L_z=4,8$ at $Re=4,000, Pm=1$. 
		We find that the stresses increase with $L_y/L_x$ quite significantly and that the saturated level of stresses for $L_y/L_x = 8$ is $\sim 40$ times larger than the case with $L_y=L_z=4$.}
\end{figure}
In the previous section, we discussed transport properties of simulations with $L_x=1,L_y=2$ and variable $L_z$. This is largely the set of runs Paper I was based on. In order to explore the sensitivity of the domain size in the `$y$' direction, we conducted two more numerical simulations with $L_y=L_z=4,8$. Large domains in both `$y$' and `$z$' directions are computationally prohibitive since, for example, a resolution of $64/L_x$ with $L_y=L_z=16$ would amount to simulating a box with $64\times 1024\times 1024$ grid points. We use the same resolution as before ($64/L_x$) but a lower Reynolds numbers, $Re=Rm=4000$. In fig. \ref{fig:alphaLyLz}, we plot the evolution of turbulent stresses for the two runs considered in this section. We find that the stress for the $L_y=L_z=4$ remains relatively similar to $L_y=2,L_z=4$ case in fig. \ref{fig:alphavsLz}  ($\alpha \sim 5\times10^{-3}$ here as opposed to $2\times10^{-3}$) but $L_y=L_z=8$ reaches higher saturated values ($\alpha \sim 2\times 10^{-1}$ that is about $50$ times larger than the $L_y=2,L_z=8$ case in fig. \ref{fig:alphavsLz}). This result is very intriguing since the $Re,Rm$ used for these runs is actually lower than the `tall' box runs so one would naively expect a lower $\alpha$ here. 

\begin{figure}
	\centering
	\includegraphics[width=0.475\textwidth]{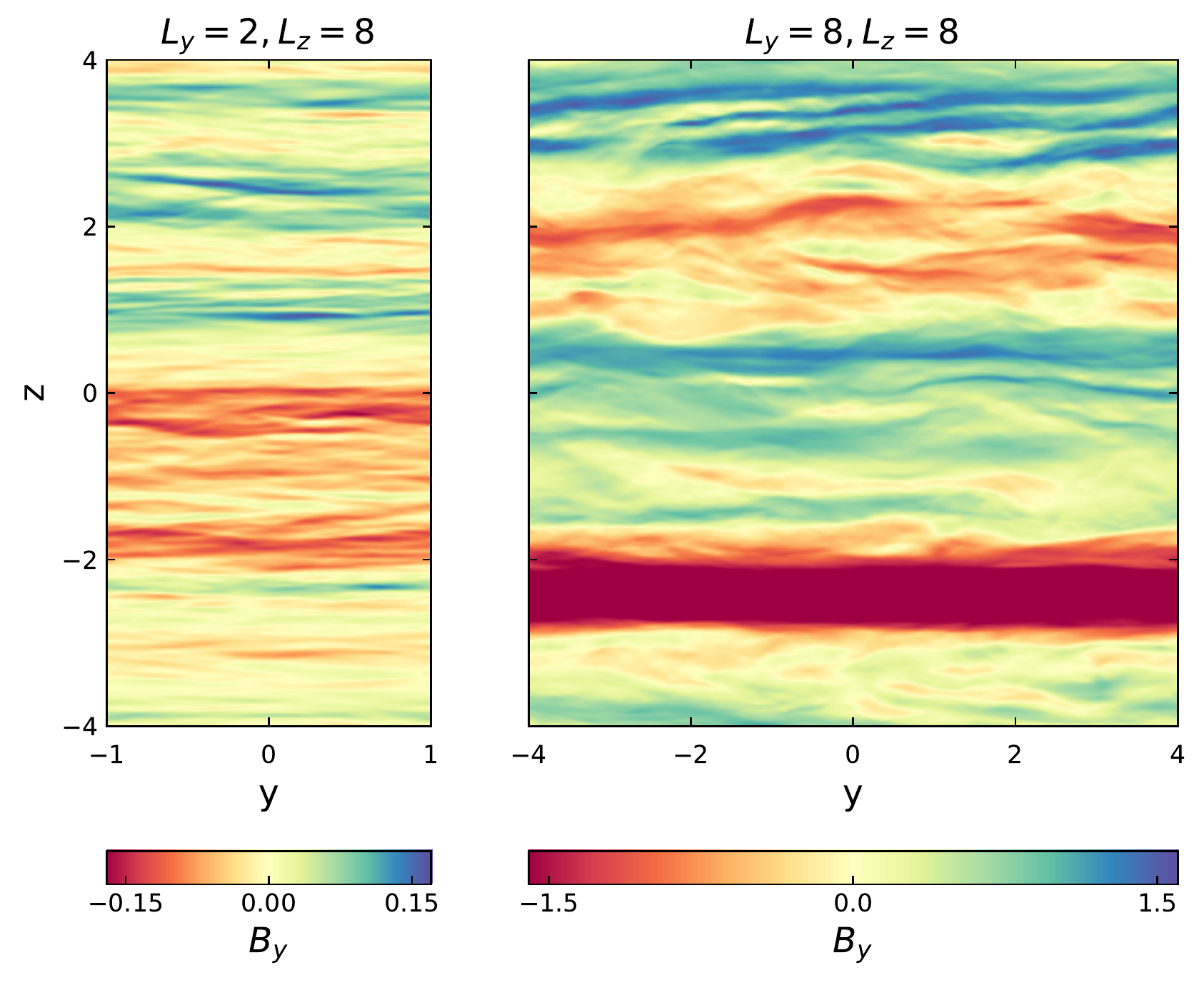}
	\label{fig:Bycomparison}
	\caption{Snapshot of $B_y$ at $t=1000S^{-1}$ averaged over `$x$'. The colorbar at the bottom indicates nearly an order of magnitude difference in magnitudes of $B_y$ for the two cases. \textit{Left}: Some hint for large scale structure exists. \textit{Right}: The extended `$yz$' domain case presents some strong bands that seem to be also responsible for $\overline{b^2}\sim\overline{B}^2$.}
\end{figure}
When averaged over the time period $250-500S^{-1}$, the extended $L_y=L_z=8$ behaves similar to the $L_y=2,L_z=8$ simulation reported in the previous section: $\lb \overline{-b_xb_y}\rb \sim \lb\overline{-b'_xb'_y}\rb$ and $\lb\overline{b^2}\rb \sim \lb\overline{b'^2}\rb$. However, in the saturated state ($\geq 500S^{-1}$), there is a significant change in behavior: the magnetic energy is dominated by the mean fields: $\lb\overline{b^2}\rb \sim \lb\overline{B}^2\rb$ but the stress is still dominated by the fluctuating component. In fig. \ref{fig:Bycomparison}, we show the snapshot of the azimuthal magnetic for the two $L_z=8$ runs: one with $L_y=2, Re=Rm=10000$ and the other with $L_y=8, Re=Rm=4000$. In both cases, the field has considerable structure in the vertical direction. A strong banded structure parallel to the $y$ axis forms in the extended $yz$ domain case while the `tall' box case has weaker bands but more of them (see also \cite{walker2017}).

\section{Conclusions} \label{sec:conclusions}
Most studies of magnetized Keplerian flows have focused on an imposed magnetic flux in a small domain. Some of the literature has focused on studying the effects of dissipation coefficients on sustained turbulence \citep{lesur2007,meheut}. Recent work has suggested that domain size might play a key role when there is no imposed flux \citep{shi2016,naumanpessah2016}. This work is a survey of the effects of large aspect ratios and dissipation coefficients on transport coefficients. We have found that:
\begin{enumerate}
	\item For fixed $L_y/L_x=2$, $\alpha$ is not sensitive to $Pm$ for $Pm < 1$ and follows a power law $\alpha \sim Pm^{1.5}$ for $Pm > 1$.
	\item For fixed $L_y/L_x=2$, the turbulent stresses $\alpha$, $\alpha_{\text{mag}}$ and $\langle -b_xb_y\rangle/\langle v_xv_y\rangle$ are all nearly insensitive to increase in $L_z/L_x$.
	\item For $L_y=L_z \gg L_x$, the saturated level of $\alpha$ increases significantly especially with $L_y=L_z=8$ where $\alpha \sim 0.2$.
\end{enumerate}
Our study highlights the importance of aspect ratio that we first pointed out in Paper I and motivates further work that might give important insights into the role of turbulence in magnetized Keplerian flows.

\section*{acknowledgments}
	The research leading to these results has received funding from the European Research Council under the European Union's Seventh Framework Programme (FP/2007-2013) under ERC grant agreement 306614.

\bibliography{general}
\bibliographystyle{mnras}


\label{lastpage}

\end{document}